\documentclass[12pt]{article}
\begin{document}

\def\bd{\begin{displaymath}}\def\ed{\end{displaymath}}
\def\be{\begin{equation}}\def\ee{\end{equation}}
\def\bea{\begin{eqnarray}}\def\eea{\end{eqnarray}}
\def\nn{\nonumber}\def\lb{\label}
 
\def\a{\alpha}\def\b{\beta}\def\c{\chi}\def\d{\delta}\def\e{\epsilon}
\def\f{\phi}\def\g{\gamma}\def\h{\theta}\def\i{\iota}\def\j{\vartheta}
\def\k{\kappa}\def\l{\lambda}\def\m{\mu}\def\n{\nu}\def\o{\omega}\def
\p{\pi}\def\q{\psi}\def\r{\rho}\def\s{\sigma}\def\t{\tau}\def\u{\upsilon}
\def\vu{\varphi}\def\w{\varpi}\def\y{\eta}\def\x{\xi}\def\z{\zeta}

\def\D{\Delta}\def\F{\Phi}\def\G{\Gamma}\def\H{\Theta}\def\L{\Lambda}
\def\O{\Omega}\def\P{\Pi}\def\Q{\Psi}\def\S{\Sigma}\def\U{\Upsilon}\def\X{\Xi}

\def\lie{{\cal L}}\def\de{\partial}\def\na{\nabla}\def\per{\times}
\def\inf{\infty}\def\id{\equiv}\def\mo{{-1}}\def\ha{{1\over 2}}
\def\qu{{1\over 4}}\def\pro{\propto}\def\app{\approx}
\def\we{\wedge}\def\di{{\rm d}}\def\Di{{\rm D}}

\def\Ei{{\rm Ei}}\def\li{{\rm li}}\def\const{{\rm const}}\def\ex{{\rm e}}
\def\arcsh{{\rm arcsinh}}\def\arcch{{\rm arccosh}}
\def\arcth{{\rm arctanh}}\def\arccth{{\rm arccoth}}
\def\diag{{\rm diag}}

\def\fe{field equations }\def\bh{black hole }\def\as{asymptotically }
\def\tran{transformations }\def\coo{coordinates }
\def\crel{commutation relations }\def\poi{Poincar\'e }
\def\des{de Sitter }\def\dpa{deformed \poi algebra }

\def\PL#1{Phys.\ Lett.\ {\bf#1}}\def\CMP#1{Commun.\ Math.\ Phys.\ {\bf#1}}
\def\PRL#1{Phys.\ Rev.\ Lett.\ {\bf#1}}\def\AP#1#2{Ann.\ Phys.\ (#1) {\bf#2}}
\def\PR#1{Phys.\ Rev.\ {\bf#1}}\def\CQG#1{Class.\ Quantum Grav.\ {\bf#1}}
\def\NP#1{Nucl.\ Phys.\ {\bf#1}}\def\GRG#1{Gen.\ Relativ.\ Grav.\ {\bf#1}}
\def\JMP#1{J.\ Math.\ Phys.\ {\bf#1}}\def\PTP#1{Prog.\ Theor.\ Phys.\ {\bf#1}}
\def\PRS#1{Proc.\ R. Soc.\ Lond.\ {\bf#1}}\def\NC#1{Nuovo Cimento {\bf#1}}
\def\JP#1{J.\ Phys.\ {\bf#1}} \def\IJMP#1{Int.\ J. Mod.\ Phys.\ {\bf #1}}
\def\MPL#1{Mod.\ Phys.\ Lett.\ {\bf #1}} \def\EL#1{Europhys.\ Lett.\ {\bf #1}}
\def\AIHP#1{Ann.\ Inst.\ H. Poincar\'e {\bf#1}}\def\PRep#1{Phys.\ Rep.\ {\bf#1}}
\def\AoP#1{Ann.\ Phys.\ {\bf#1}}
\def\grq#1{{\tt gr-qc/\-#1}}\def\hep#1{{\tt hep-th/\-#1}}

\def\den{\left(1-{P_0\over\k}\right)}\def\dem{\left(1-{\y_0\over\k}\right)}
\def\by{\bar\y}

\begin{titlepage}
\vspace{.3cm}
\begin{center}
\renewcommand{\thefootnote}{\fnsymbol{footnote}}
{\Large \bf Two-dimensional gravity with an invariant energy scale}
\vfill
{\large \bf {S.~Mignemi\footnote{email: smignemi@unica.it}}}\\
\renewcommand{\thefootnote}{\arabic{footnote}}
\setcounter{footnote}{0}
\vfill
{\small
  Dipartimento di Matematica, Universit\`a di Cagliari,\\
Via Ospedale 72, 09123 Cagliari, Italy\\
\vspace*{0.4cm}
 INFN, Sezione di Cagliari\\
}
\end{center}
\vfill
\centerline{\bf Abstract}
\vfill

We investigate the gauging of a two-dimensional deformation of the
Poin\-car\'e algebra, which accounts for the existence
of an invariant energy scale. The model describes 2D dilaton gravity with
torsion. We obtain explicit solutions of the field equations and
discuss their physical properties.

\vfill
\end{titlepage}

Investigation of quantum gravity and string theory seems to
indicate the existence of a fundamental length scale of the order
of the Planck length \cite{Ga}, that may also give rise to observable
effects \cite{obs}. However, a fundamental
frame-independent length (or equivalently, energy scale)
cannot be introduced without modifying
special relativity, since it would break the invariance of the
theory under the \poi group \cite{AC}.
Recently, it was observed by Magueijo and Smolin (MS) \cite{MS} that it
is nevertheless possible to preserve the invariance under the subgroup
of Lorentz transformations,
assuming that its action on momentum space is non-linear.
As remarked in \cite{KN}, this proposal can be interpreted as a
special case of a larger class of deformations of
the Poincar\'e algebra which were introduced in \cite{dpa}.
The effect of these deformations would be appreciable only for energy
scales of the order of the Planck energy, while for smaller scales
one would recover special relativity.

An interesting problem is how to include gravity in this framework.
One may hope that the modification of the short-distance behaviour
of the theory induced by the existence of a minimal length could
avoid the singularities which affect general relativity.
Of course, the most straightforward way to introduce gravity
is by gauging the \dpa of Ref. \cite{MS}. This algebra can be considered
as a special case of non-linear algebra. The gauge theory of non-linear
algebras has been studied some time ago \cite{Ike,van}, but
unfortunately a suitable action for these models has been obtained only
in two dimensions \cite{Ike}\footnote{ An alternative formalism based on
Poisson sigma-models was introduced in \cite{stro}.}.

In this letter, we apply the formalism of \cite{Ike} to the study
of the two-dimensional version of the MS algebra. We obtain a model
of 2D gravity that modifies those based on the \poi algebra \cite{Jack}.
An interesting consequence of the breaking of the \poi invariance is
that non-trivial torsion is present in the theory. This seems to
be a essential feature of models of this kind.

\bigbreak
The non-linear \dpa of \cite{MS} is given in two dimensions by the \crel
\be\lb{algebra}
[P_a,P_b]=0,\quad [J,P_0]=\den P_1,\quad
[J,P_1]=P_0-{P_1^2\over\k},
\ee
where $P_a$ are the generators of translations and $J$ that of
boosts and $a=0,1$. Tangent space indices are lowered and raised
by the tensor $h_{ab}=\diag(-1,1)$. We also make use of the antisymmetric tensor
$\e_{ab}$, with $\e_{01}=1$. The deformation parameter $\k$ has the dimension
of a mass and can be identified with the inverse of the Planck length.
The algebra admits a Casimir invariant
\be\lb{Casimir}
C={P_1^2-P_0^2\over\den^2}.
\ee
In the following it will be useful to denote the generators of the algebra
as $T_A$, where $ A=0,1,2$ and $T_a=P_a$, $T_2=J$.

In order to construct a gauge theory for this algebra, we adopt
the formalism of Ikeda \cite{Ike}. Given an algebra with \crel
$[T_A,T_B]=W_{AB}(T)$, one introduces gauge fields
$A^A$ and a coadjoint multiplet of scalar fields $\y_A$, which
under infinitesimal \tran of parameter $\x^A$ transform as
\bea\lb{trans}
&&\d A^A=\di\x^A+U^A_{BC}(\y)A^B\x^C,\cr
&&\d\y_A=-W_{AB}(\y)\x^B,
\eea
where $U^A_{BC}$ and $W_{AB}$ are functions of the fields $\y$,
which satisfy
\bea
U^A_{BC}={\de W_{BC}\over\de\y_A}.
\eea
One can then define the covariant derivative of the scalar
multiplet
\be
\Di\y_A=\di\y_A+W_{AB}A^B,
\ee
and the curvature of the gauge fields
\be
F^A=\di A^A+U^A_{BC}A^B\we A^C.
\ee
In two dimensions, a gauge invariant lagrangian density can be defined
as \cite{Ike}
\be\lb{lag}
L=\y_A F^A+(W_{BC}-\y_AU^A_{BC})A^B\we A^C,
\ee
and generates the \fe
\be\lb{feq}
\Di\y^A=0,\qquad F^A=0.
\ee
In our case, the functions $W_{AB}$ can be deduced from the algebra
(\ref{algebra}).

A theory of gravity can now be defined analogously to \cite{Jack},
by identifying $A^a$ with the zweibeins $e^a$ and
$A^2$ with the spin connection $\o$. It follows that $F^2=R$ and
$F^a=T^a-{\o\over\k}\we(\y_1e^a+\y_be^b\d^a_1)$, where $R=\di\o$ is
the curvature and $T^a=\di e^a+\e^a_{\ b}\,\o\we e^b$ the torsion,
and the lagrangian (\ref{lag}) takes the form
\be\lb{lag'}
L=\y_aT^a+\y_2R+{\y_1\over\k}\y_a\,\o\we e^a.
\ee
Clearly, the last term in (\ref{lag'}) breaks the \poi invariance.

The \fe (\ref{feq}) read explicitly
\bea
&&\di\y_0-\o\left(\y_1-{\y_1\y_0\over\k}\right)=0,\lb{fey1}\\
&&\di\y_1-\o\left(\y_0-{\y_1^2\over\k}\right)=0,\lb{fey2}\\
&&\di\y_2+\left(\y_1-{\y_1\y_0\over\k}\right)e^0+\left(\y_0-
{\y_1^2\over\k}\right)e^1=0,\lb{fey3}\\
&&\di e^0+\o\we\left(-{\y_1\over\k}e^0+e^1\right)=0,\lb{fea1}\\
&&\di e^1+\o\we\left[\left(1-{\y_0\over\k}\right)e^0-
{2\y_1\over\k}e^1\right]=0,\lb{fea2}\\
&&\di\o=0.\lb{fea3}
\eea
In spite of their complexity, they can be solved generalizing a
method introduced by Solodukhin \cite{Sol} for a different 2D
gravity model. We give here only a short account of the main steps
necessary for obtaining the solution.

First define new fields $\bar\y_a=\dem^\mo\y_a$,
which satisfy the relations
$\di\by_a=\e_a^{\ b}\by_b$ and $\di(\by_a\by^a)=0$. Hence
$\by^2=\by_a\by^a$ is a constant of the motion, $\by^2=-a^2$, say,
the negative sign corresponding to a positive "mass" squared
(cf.\ eq.\ (\ref{Casimir})).
The previous relations suggest to define a variable $\h$ such that
$\by_0=a\cosh\h$, $\by_1=a\sinh\h$, and therefore
$\by_a\e^{ab}\di\by_b=-a^2\di\h$. But the \fe (\ref{fey1},\ref{fey2}),
yield $\by_a\e^{ab}\di\by_b=\o\by_a\by^a=-a^2\o$, and hence
$\o=\di\h$, in accordance with (\ref{fea3}).
Moreover, combining (\ref{fey1},\ref{fey2}) with
(\ref{fea1},\ref{fea2}), it is easy to see that
$\di\!\left[\dem\y_ae^a\right]=0$, which allows one to define a new
variable $\f$, such that $\y_ae^a=\dem^\mo\di\f$.

Finally, writing (\ref{fey3}) as
$\e^a_{\ b}\y_ae^b=\di\y_2-{\y_1\over\k}\y_ae^a$,
and comparing with the previous equation,
one can solve for $e^0$ and $e^1$.
The result is
\bea
&&e^0=-{\D\over a}\left[\sinh\h\left(\di\y_2-{a\over\k}
\sinh\h\ \di\f\right)+\D\cosh\h\ \di\f\right],\cr
&&e^1={\D\over a}\left[\cosh\h\left(\di\y_2-{a\over\k}
\sinh\h\ \di\f\right)+\D\sinh\h\ \di\f\right],
\eea
where $\D=\dem^\mo=1+{a\over\k}\cosh\h$.

One is still free to choose a gauge. The most interesting choices
are $\h=0$ and $\di\h=\D\di\f$.
The first choice leads to flat space with vanishing torsion. In the
second case,
\be
e^0=-{\D\over a}(\sinh\h\di\q+\cosh\h\di\h),\qquad
e^1={\D\over a}(\cosh\h\di\q+\sinh\h\di\h),
\ee
where we have defined a new coordinate
$\q=\y_2-\log\D$,
and the components of the torsion are
\be
T^0=-{1\over\k}\sinh^2\h\ \di\h\we\di\q,\qquad
T^1={1\over\k}\sinh\h\cosh\h\ \di\h\we\di\q,
\ee
or, in orthonormal coordinates,
\be
T^0_{01}=-{\sinh^2\h\over\k\,\D^2},\qquad
T^1_{01}={\sinh\h\cosh\h\over\k\,\D^2},
\ee
while the curvature vanishes identically, due to (\ref{fea3}).
This solution presents no singularities, since $\D>0$ everywhere.
The coordinate $\h$ being timelike, the solution can be
interpreted as a cosmological one.

One may also define a spacetime metric as
\be
ds^2\id h_{ab}e^ae^b={1\over a^2}\left(1+{a\over\k}\cosh\h\right)
(-\di\h^2+\di\q^2).
\ee
However, this quantity is not gauge-invariant, due to the nonlinear
transformation properties (\ref{trans}) of the zweibeins. One may
still try to define a gauge-invariant metric as a more general
quadratic form in the zweibeins with $\y$-dependent coefficients,
but we were not able to find a suitable expression (see however
\cite{GKSV}). A possible interpretation of
the gauge dependence of the metric in the context of models
based on \cite{AC,MS} is that different sub-Planckian observers
"see" a different spacetime metric depending on their momentum.

The \fe also admit singular solutions for $\by^2=a^2>0$.
In this case one has
\be\lb{sol}
e^0=-{\U\over a}(\cosh\h\di\q+\sinh\h\di\h),\qquad
e^1={\U\over a}(\sinh\h\di\q+\cosh\h\di\h),
\ee
with $\U=1+{a\over\k}\sinh\h$, and
\be
T^0=-{1\over\k}\cosh^2\h\ \di\h\we\di\q,\qquad
T^1={1\over\k}\sinh\h\cosh\h\ \di\h\we\di\q.
\ee
While the curvature is zero everywhere, the invariants built
with the torsion diverge when $\U=0$ (i.e. $\h=\h_0=\arcsh(-\k/a$)),
and hence the solution is singular there. The coordinate $\h$ is now
spacelike and hence the solution may be interpreted
as a singular black-hole spacetime. In fact, one may define
as before a (not gauge invariant) line element
\be ds^2\id h_{ab}e^ae^b={1\over a^2}\left(1+{a\over\k}\sinh\h\right)
(-\di\q^2+\di\h^2),
\ee
which exhibits a horizon at $\h=\h_0$.
\bigbreak

We have shown that it is possible to construct a theory of gravity in
two dimensions based on the MS algebra. Besides flat space, the model
admits regular solutions of cosmological type, but no regular solution
of \bh type.
Of course, this is the simplest model one can imagine (its
Poincar\'e-invariant limit possesses only flat solutions), and one may
consider models based on different deformations of 2D \poi or \des
algebras, which may show more attractive features. This topic is
currently being investigated \cite{GKSV}.
Also, it would be interesting to include matter in order to
obtain more physical insight on the properties of the theory.

At first sight, it seems that the introduction of the new invariant
parameter $\k$ affects the global properties of the solutions, rather
than their short-distance behavior. However, one must remember
that it is not possible to construct a gauge-invariant metric, and
therefore the geometry of the theory must be interpreted only in terms
of the zweibein and the spin connection. This requires a more careful
analysis.
Perhaps the introduction of non-commuting spacetime \coo is
necessary for a better understanding of this point \cite{dpa}.

The most interesting development of our results would be of course
their extension to higher dimensions.
This would require the definition of an action suitable for
non-linear gauge theories in $D>2$, which is not known at
present. An important point however is that all models of this kind
imply the presence of non-trivial torsion, which could also affect the
coupling of matter.
\bigskip

{\bf Acknowledgments.}
I wish to thank T. Strobl for some interesting comments on a
previous version of this work.

\end{document}